\renewcommand{\theequation}{\arabic{section}.\arabic{equation}}
\documentclass{ijsipaper}
\ifprodtf \else \usepackage{latexsym}\fi
\newcommand\black{\ensuremath{\blacktriangleright}}
\newcommand\white{\ensuremath{\vartriangleright}}
\newif\ifamsfontsloaded
\ifprodtf
  \newcommand\whbl{\white\kern-.1em--\kern-.1em\black}
  \newcommand\blwh{\black\kern-.1em--\kern-.1em\white}
  \newcommand\blbl{\black\kern-.1em--\kern-.1em\black}
  \newcommand\whwh{\white\kern-.1em--\kern-.1em\white}
  \amsfontsloadedtrue
\else
  \checkfont{msam10}
  \iffontfound
    \IfFileExists{amssymb.sty}
      {\usepackage{amssymb}\amsfontsloadedtrue
       \newcommand\whbl{\white\kern-.125em--\kern-.125em\black}%
       \newcommand\blwh{\black\kern-.125em--\kern-.125em\white}%
       \newcommand\blbl{\black\kern-.125em--\kern-.125em\black}%
       \newcommand\whwh{\white\kern-.125em--\kern-.125em\white}}
      {}
  \fi
\fi

\usepackage[matrix,frame,arrow]{xypic}%    Q-circuit version 1.2
\usepackage[matrix,frame,arrow]{xy}
\usepackage{amsmath}

\newcommand{\qw}[1][-1]{\ar @{-} [0,#1]}
\newcommand{\gate}[1]{*{\xy *+<.6em>{#1};p\save+LU;+RU **\dir{-}\restore\save+RU;+RD **\dir{-}\restore\save+RD;+LD **\dir{-}\restore\POS+LD;+LU **\dir{-}\endxy} \qw}
\newcommand{\measureD}[1]{*{\xy*+=+<.5em>{\vphantom{\rule{0em}{.1em}#1}}*\cir{r_l};p\save*!R{#1} \restore\save+UC;+UC-<.5em,0em>*!R{\hphantom{#1}}+L **\dir{-} \restore\save+DC;+DC-<.5em,0em>*!R{\hphantom{#1}}+L **\dir{-} \restore\POS+UC-<.5em,0em>*!R{\hphantom{#1}}+L;+DC-<.5em,0em>*!R{\hphantom{#1}}+L **\dir{-} \endxy} \qw}
\newcommand{\multigate}[2]{*+<1em,.9em>{\hphantom{#2}} \qw \POS[0,0].[#1,0];p !C *{#2},p \save+LU;+RU **\dir{-}\restore\save+RU;+RD **\dir{-}\restore\save+RD;+LD **\dir{-}\restore\save+LD;+LU **\dir{-}\restore}
\newcommand{\ghost}[1]{*+<1em,.9em>{\hphantom{#1}} \qw}
\newcommand{\Qcircuit}[1][0em]{\xymatrix @*=<#1>} % @*[o]
\newcommand{\pureghost}[1]{*+<1em,.9em>{\hphantom{#1}}}
\newcommand{\multiprepareC}[2]{*+<1em,.9em>{\hphantom{#2}}\save[0,0].[#1,0];p\save !C
  *{#2},p+RU+<0em,0em>;+LU+<+.8em,0em> **\dir{-}\restore\save +RD;+RU **\dir{-}\restore\save
  +RD;+LD+<.8em,0em> **\dir{-} \restore\save +LD+<0em,.8em>;+LU-<0em,.8em> **\dir{-} \restore \POS
  !UL*!UL{\cir<.9em>{u_r}};!DL*!DL{\cir<.9em>{l_u}}\restore}
\newcommand{\prepareC}[1]{*{\xy*+=+<.5em>{\vphantom{#1\rule{0em}{.1em}}}*\cir{l^r};p\save*!L{#1} \restore\save+UC;+UC+<.5em,0em>*!L{\hphantom{#1}}+R **\dir{-} \restore\save+DC;+DC+<.5em,0em>*!L{\hphantom{#1}}+R **\dir{-} \restore\POS+UC+<.5em,0em>*!L{\hphantom{#1}}+R;+DC+<.5em,0em>*!L{\hphantom{#1}}+R **\dir{-} \endxy}}
\newcommand{\poloFantasmaCn}[1]{{{}^{#1}_{\phantom{#1}}}}

\newcommand{\Tr}{\operatorname{Tr}}

\newcommand{\cat}[1]{\mathbf{#1}}
\newcommand{\set}[1]{\mathsf{#1}}
\newcommand{\map}[1]{\mathcal{#1}}

\newcommand{\type}[1]{\mathrm{#1}}
\newcommand{\rA}{\type{A}}
\newcommand{\rB}{\type{B}}
\newcommand{\rC}{\type{C}}
\newcommand{\rD}{\type{D}}
\newcommand{\rE}{\type{E}}

\newcommand{\rR}{\type{R}}
\newcommand{\rI}{\type{I}}

\newcommand{\Pure}{{\mathsf{Pure}}}

\def\Proof{\medskip\par\noindent{\bf Proof. }}

\def\qed{$\square$}
\def\>{\rangle}
\def\<{\langle}

\newtheorem{definition}{Definition}[section]
\newtheorem{proposition}{Proposition}[section]
\newtheorem{axiom}{Axiom}%[section]

\title{Distinguishability and copiability of programs  in general process theories}

\author[G. Chiribella]
    {Giulio Chiribella\vspace*{2mm}\\
     Center for Quantum Information, Institute for Interdisciplinary Information Sciences, \\
     Tsinghua University, Beijing 100084, China}

\pubyear{2014} %
\pubmonth{January} %
\pubnumber{2} %
\volume{1} %

\sponsor{Institute of Software, Chinese Academy of Sciences} %
\receiveddate{2009-03-08} %
\reviseddate{2009-10-10} %
\accepteddate{2009-10-10} %
\onlinepulishdate{2009-12-12} %
\pagerange{1--14}
\begin{document}

\setcounter{page}{1}

\label{firstpage}

\maketitle
%\makesponsor %
%\makecorrespond %
%\makeusefuldate %

\begin{abstract}
We propose a notion of state distinguishability that does not refer to probabilities, but rather to the ability of a set of states to serve as programs for a desired set of gates.  Using this notion, we reconstruct the structural features of the task of state discrimination, such as the equivalence with cloning and the  impossibility to extract information from two non-distinguishable pure states without causing a disturbance.  All these features express intrinsic links among operational tasks, which are valid independently of the particular theory under consideration.  
% the   
 % cryptographic version of the No Information Without Disturbance principle.
\end{abstract}


%\makepaperinfo\normalsize\parindent=7mm

\section{Introduction}  

Quantum information science is making rapid progress towards the realization of a new generation of technologies which promises to revolutionize communication, sensing, and computation \cite{Nie01,Mer07,Wil13}.     At the same time, it is  also revolutionizing   the very notion of information, questioning the preeminence of classical information theory as \emph{the} canonical theory of information.  
% Indeed, while  classical theory provides an accurate modelling of information-processing at the macroscopic scale, it clearly fails to do so at the microscopic scale.    

The sharp contrast between the familiar world  of classical information and the exotic features of quantum information   quickly leads to  questions about alternative theories of information  beyond quantum theory:  after all, we expected the world to be classical and discovered that it was quantum---what if one day we were to discover another, more fundamental theory of physics, which is neither classical nor quantum?  Should we revise the conceptual framework of information theory once more?  This type of questions motivated the investigation of a larger class of theories, broadly known as \emph{general probabilistic theories (GPTs)}   \cite{Har01,DAr06,Bar07,Lei07,Spe07,Chi10,DAr10,Chi11,Har11,Dak11,Bar11,Mas11,Bar12,Mas13}.    GPTs provide a neutral framework  for analyzing high level features of information-processing  protocols,  independently of the particular laws that govern the hardware level.   
 The GPT framework has been used for characterizing quantum theory in terms of operational axioms \cite{Har01,Chi11,Har11,Dak11,Mas11,Mas13} and for establishing direct links among  information-theoretic  protocols \cite{Bar07,Lei07,Chi10,Bar11,Bar12}.  
 This analysis has the merit of identifying  structures in the family of quantum protocols and of highlighting general ``laws   of  information" that are   independent  of the specific theory under consideration.     An example of such a law is the equivalence between cloning and distinguishability, stating that different pieces of information can be replicated  if and only if they are distinguishable from one another.

In the spirit of identifying  theory-independent ``laws of information" and establishing direct links among protocols  one can also go one step further. Instead of probabilistic theories, one can consider more primitive theories that only describe operations, without assigning probabilities to the random events that may be generated by these operations.  The study of these theories, now known as \emph{process theories},  has been pioneered by Abramsky and Coecke \cite{Abr04,Coe06,Abr08,Coe10}      and  has  been extensively employed for the reconstruction of  quantum protocols over the past ten years
\cite{Abr04,Abr09,Hor11,Coe12,Ran14}.    

While a full characterization of quantum theory in the language of process theories seems to be still far away, it is stimulating to ask how far the probability-free  approach can go.    At the level of principles this is an important question, because it aims at drawing the line between those aspects of information that are defined only in terms of operations (and therefore can  be mechanized)  and those that rely on  the subjective expectations of an  agent.        
This paper provides a contribution in this direction, presenting a probability-free treatment of the relations between distinguishability, cloning, and programming of gates.

\section{Process theories}  

The  framework of \emph{process theories}  \cite{Abr04,Coe06,Coe10}  is  based on  (strict) symmetric monoidal categories (SMC) \cite{Awo10}.   Nevertheless,  no prior knowledge of category theory is required to understand its basic features: the basic categorical facts are encoded in a  graphical language \cite{Sel11} which is identical to the familiar languages of quantum circuits and of classical Boolean circuits.   The role of SMCs    is just  to provide  the mathematical foundations of such graphical language.    

\subsection{An abstract circuit model}

A process theory describes circuits that transform input data into  output data, or,  in the terminology of physics, circuits that transform input systems into output systems.   Mathematically, a ``process theory" is an SMC, here denoted by $\cat C$.  
The objects in the category, denoted by $ |\cat C|$, represent different data types  (a.k.a. different systems, in the physics terminology).      The morphisms in the category represent the different gates (a.k.a. different physical processes) that transform an input system into an output system. A gate of type $\rA\to \rB$ will be represented as 
\begin{align*}
\begin{aligned}  
\Qcircuit @C=.5em @R=0em @!R {     & \qw \poloFantasmaCn{\rA}  &\gate{\map G}   &   \qw \poloFantasmaCn{\rB}   & \qw} 
\end{aligned}    \quad .
\end{align*}     
The set of all gates of type $\rA\to \rB$ will be denoted  by $\cat C (\rA,\rB)$.  
For two gates $\map G\in \cat C (\rA,\rB)$ and $\map H \in\cat C(\rB,\rC)$, we denote the sequential composition as $ \map G  ;  \map H   \in \cat C  (\rA,\rC)$  and represent it graphically as
\begin{align*}
\begin{aligned}  \Qcircuit @C=.5em @R=0em @!R {     
& \qw \poloFantasmaCn{\rA}  &\gate{\map G}   &   \qw \poloFantasmaCn{\rB}   & \gate{\map H}    &   \qw \poloFantasmaCn{\rC}  &\qw } 
\end{aligned} \, .
\end{align*}     
The identity gate on $\rA$, denoted by $\map I_\rA$, will be represented equivalently as
\begin{align*}
\begin{aligned}  
\Qcircuit @C=.5em @R=0em @!R {     & \qw \poloFantasmaCn{\rA}  &\gate{\map I}   &   \qw \poloFantasmaCn{\rA}   & \qw} 
\end{aligned} \quad{\rm and }\quad  
\begin{aligned}  
\Qcircuit @C=.5em @R=0em @!R {     &\qw & \qw \poloFantasmaCn{\rA}  &\qw   &   \qw      \qw}  
\end{aligned}    \quad .
\end{align*}     
%The identity on the trivial system will be denoted by ${\sf 1}$. 
A gate $\map U  :  \rA\to \rB$ is  \emph{reversible}    (or equivalently, is an \emph{isomorphism}) iff there exists another gate $\map U^{-1}  :  \rB\to \rA$ such that   
\begin{align*}
\begin{aligned}  \Qcircuit @C=.5em @R=0em @!R {     
& \qw \poloFantasmaCn{\rA}  &\gate{\map U}   &   \qw \poloFantasmaCn{\rB}   & \gate{\map U^{-1}}    &   \qw \poloFantasmaCn{\rA}  &\qw      &  & \quad & = &\quad &         &\qw & \qw \poloFantasmaCn{\rA}  &\qw   &   \qw      \qw}
\end{aligned}   \qquad  {\rm and  \qquad }  \begin{aligned}  \Qcircuit @C=.5em @R=0em @!R {     
& \qw \poloFantasmaCn{\rB}  &\gate{\map U^{-1}}   &   \qw \poloFantasmaCn{\rA}   & \gate{\map U}    &   \qw \poloFantasmaCn{\rB}  &\qw      &  & \quad & = &\quad &         &\qw & \qw \poloFantasmaCn{\rB}  &\qw   &   \qw      \qw}
\end{aligned}\quad .
\end{align*}     
If there exists a reversible gate of type $\rA\to \rB$, the systems $\rA$ and $\rB$ are called \emph{isomorphic}, denoted as $\rA\simeq \rB$. 
  
When two  systems  $\rA$ and $\rB$  are considered together, we denote their tensor as  $\rA\otimes \rB$.  The absence of relevant data is represented by  the monoidal unit, denoted by $\rI$.         
When two gates $\map A\in \cat C(\rA,\rA')$ and $\map B\in\cat C (\rB,\rB')$ operate in parallel, their action is described by the tensor product gate $\map A\otimes \map B\in\cat C(\rA\otimes \rB,\rA'\otimes \rB')$, graphically represented as 
\begin{align*}
\begin{aligned}  \Qcircuit @C=.5em @R=0em @!R {     & \qw \poloFantasmaCn{\rA}  &\gate{\map A}   &   \qw \poloFantasmaCn{\rA'}   & \qw\\   
\\
& \qw \poloFantasmaCn{\rB}  &\gate{\map B}   &   \qw \poloFantasmaCn{\rB'}   & \qw} 
\end{aligned}     \quad . 
\end{align*}     

Motivated by the physical interpretation,   a gate   $\rho$ of type  $\rI\to \rA$ will be called a \emph{state of system $\rA$} and will be represented as 
\begin{align}
\begin{aligned}  
\Qcircuit @C=.5em @R=0em @!R {     &\prepareC {\rho}  & \qw \poloFantasmaCn{\rA}    & \qw} 
\end{aligned}     
:= 
\begin{aligned}  
\Qcircuit @C=.5em @R=0em @!R {     & \qw \poloFantasmaCn{\rI}  &\gate{\rho}   &   \qw \poloFantasmaCn{\rA}   & \qw} 
\end{aligned}    \quad .
\end{align}     
A gate $a$ of type $\rA\to \rI$ will be called an \emph{effect on system $\rA$} and will be represented as 
\begin{align}
\begin{aligned}  
\Qcircuit @C=.5em @R=0em @!R {  & \qw \poloFantasmaCn{\rA}    & \measureD a} 
\end{aligned}     
:= 
\begin{aligned}  
\Qcircuit @C=.5em @R=0em @!R {     & \qw \poloFantasmaCn{\rA}  &\gate{a}   &   \qw \poloFantasmaCn{\rI}   & \qw} 
\end{aligned}    \quad .
\end{align}     
 A gate $s$ of  type $\rI\to \rI$ will be called a \emph{scalar} and will be sometimes represented ``out of the box", as   
\begin{align}\label{outofbox}
s:= 
\begin{aligned}  
\Qcircuit @C=.5em @R=0em @!R {     & \qw \poloFantasmaCn{\rI}  &\gate{s}   &   \qw \poloFantasmaCn{\rI}   & \qw} 
\end{aligned}    \quad .
\end{align}     
We denote the identity gate on system $\rI$ as $1$.   Recall that scalars in an SMC form a commutative monoid \cite{Kel80}, with $      s  ;   { 1}   =   { 1}  ;  s  =  s$ for every scalar $s\in\cat C(\rI,\rI)$.

\subsection{Causality}  

An important requirement for a physical  theory is \emph{causality} \cite{Chi10,Chi11}.  Informally, causality is the requirement that  information in a circuit flows from the input to the output, and not vice-versa.  
  In the categorical language, causality is formulated as  \emph{terminality of the tensor unit}   \cite{Lal12,Coe14}: 
    \begin{axiom}[Causality]\label{ax:caus}
For every system  $\rA  \in  |\cat C|$ there exists one and only one  gate of type $\rA\to \rI$, called the \emph{trace on $\rA$}, denoted by $\Tr_\rA$, and represented as    
$\begin{aligned}  
\Qcircuit @C=.5em @R=0em @!R {     & \qw \poloFantasmaCn{\rA}    & \measureD{\Tr}} 
\end{aligned}$.  
  \end{axiom}
  \medskip 
  
Note that, by definition, one has $\Tr_\rI  =  1$.   Note also that, by definition, every state   $ \rho   \in   \cat C(\rI,\rA)$  is normalized as  
\begin{align}\label{statenorm}
\begin{aligned}  
\Qcircuit @C=.5em @R=0em @!R {     &\prepareC {\rho}  & \qw \poloFantasmaCn{\rA}    & \measureD{\Tr}} 
\end{aligned}        \quad =  \quad 1   
\end{align} 
and, more generally, every gate  $\map G \in   \cat C(\rA,\rB) $ is normalized as 
\begin{align}\label{gatenorm}
\begin{aligned}  
\Qcircuit @C=.5em @R=0em @!R {       & \qw \poloFantasmaCn{\rA}       &\gate {\map G}  & \qw \poloFantasmaCn{\rB}    & \measureD{\Tr}} 
\end{aligned}        \quad =  \quad     \begin{aligned}  
\Qcircuit @C=.5em @R=0em @!R {     & \qw \poloFantasmaCn{\rA}    & \measureD{\Tr}} 
\end{aligned}   \, .
\end{align} 
\medskip

\subsection{Marginals and extensions}

Thanks to Causality, one can define marginal states:  
\begin{definition} 
The \emph{marginal on $\rA$} of a state $\sigma \in \cat C (\rI,\rA\otimes \rB)$   is the state   $\rho  \in  \cat C(\rI,\rA)$ defined by  
\begin{equation*}
\begin{aligned}  \Qcircuit @C=.5em @R=0em @!R 
{ & \prepareC{\rho}    &  \qw  \poloFantasmaCn{\rA}  &\qw  } 
\end{aligned} 
:=
\begin{aligned}  \Qcircuit @C=.5em @R=0em @!R { & \multiprepareC{2}{\sigma}    & \qw \poloFantasmaCn{\rA}  &  \qw \\
& \pureghost{\sigma}   &&\\
 & \pureghost{\sigma}    & \qw \poloFantasmaCn{\rB}  &  \measureD{\Tr}} 
\end{aligned}   \quad .   
\end{equation*}
When the above equation holds, we say that $\sigma$ is an \emph{extension of $\rA$ to the context $\rB$}.
 \end{definition}

\medskip  

The same definition can be put forward for general gates:   
\begin{definition} 
  The marginal on system $\rA'$ of a gate $\map H  \in  \cat C(\rA,\rA'\otimes \rB)$   is the gate   $\map G  \in  \cat C(\rA,\rA')$ defined by  
\begin{equation*}
\begin{aligned}  \Qcircuit @C=.5em @R=0em @!R 
{   & \qw  \poloFantasmaCn{\rA}    & \gate{  \map G }    &  \qw  \poloFantasmaCn{\rA'}  &\qw  } 
\end{aligned} 
:=
\begin{aligned}  \Qcircuit @C=.5em @R=0em @!R { &  \qw  \poloFantasmaCn{\rA}  &     \multigate{2}{\map H}    & \qw \poloFantasmaCn{\rA'}  &  \qw \\
  && \pureghost{\map H}   &&\\
 && \pureghost{\map H}    & \qw \poloFantasmaCn{\rB}  &  \measureD{\Tr}} 
\end{aligned}   \quad .   
\end{equation*}
When the above equation holds, we say that $\map H$ is an \emph{extension of $\map G$ to the context $\rB$}.
 \end{definition}

\subsection{Pure states and pure gates}  

Pure states are an essential concept both in physics and computer science.  Traditionally, they are defined as states that cannot be obtained by randomizing the preparation of the system---equivalently, states that cannot be decomposed as a convex combination of other states.     
Here, however, we did not introduce any notion of convex combination.   An expression like 
\begin{align*}
\begin{aligned}  \Qcircuit @C=.5em @R=0em @!R 
{ & \prepareC{\rho}    &  \qw  \poloFantasmaCn{\rA}  &\qw  } 
\end{aligned}   
 \quad =   \quad   
  p   \begin{aligned}  \Qcircuit @C=.5em @R=0em @!R 
{ & \prepareC{\rho_0}    &  \qw  \poloFantasmaCn{\rA}  &\qw  } 
\end{aligned}    \, + \,  (1-p)  
\begin{aligned}  \Qcircuit @C=.5em @R=0em @!R 
{ & \prepareC{\rho_1}    &  \qw  \poloFantasmaCn{\rA}  &\qw  } 
\end{aligned}     \, ,      \qquad p\in  \cat C(\rI,\rI)     
 \end{align*} 
is not legal in our language, because there is no notion of ``sum of states" and no notion of ``difference of two scalars".    

In order to introduce pure states in the framework, there are a few different options:  First, one could introduce probabilities, as it was done  in  \cite{Chi10}. In this way, the gates inherit a structure of vector space over the real numbers.    An other option is to \emph{assume} that there is a distinguished subset of states and gates that are nominally regarded as  ``pure". This approach was followed by Coecke \cite{Coe08} and Coecke-Perdrix \cite{Per10}, who defined the category of \emph{pure processes}  as a monoidal subcategory of $\cat C$.  
In this paper we will follow a third option, in which pure states and pure gates are defined only in terms of the circuit structure.    This approach is at the basis of the construction of \emph{categorical purification},  recently put forward by the author  \cite{Chi14a,Chi14b}.    In this construction, the pure states are defined as follows  
   
\begin{definition}\label{def:purstate} 
A state $\alpha  \in  \cat C(\rI,\rA)$ is \emph{pure}   iff it has only trivial extensions, that is, iff for every system $\rB\in |\cat C|$ and for every state $\sigma\in  \cat C(\rI,\rA\otimes\rB)$  one has the implication
\begin{equation*}
\begin{aligned}  \Qcircuit @C=.5em @R=0em @!R { & \multiprepareC{2}{\sigma}    & \qw \poloFantasmaCn{\rA}  &  \qw \\
& \pureghost{\sigma}   &&\\
 & \pureghost{\sigma}    & \qw \poloFantasmaCn{\rB}  &  \measureD{\Tr}} 
\end{aligned}    
\quad =  \begin{aligned}  \Qcircuit @C=.5em @R=0em @!R 
{ & \prepareC{\alpha}    &  \qw  \poloFantasmaCn{\rA}  &\qw  } 
\end{aligned} 
   \qquad  \Longrightarrow \qquad      \exists  \beta\in  \cat C (\rI,\rB)  :  \quad  
   \begin{aligned}  \Qcircuit @C=.5em @R=0em @!R { & \multiprepareC{2}{\sigma}    & \qw \poloFantasmaCn{\rA}  &  \qw \\
& \pureghost{\sigma}   &&\\
 & \pureghost{\sigma}    & \qw \poloFantasmaCn{\rB}  &  \qw} 
\end{aligned}    \quad  =    \begin{aligned}  \Qcircuit @C=.5em @R=0em @!R 
{ & \prepareC{\alpha}    &  \qw  \poloFantasmaCn{\rA}  &\qw \\
 & \prepareC{\beta}    &  \qw  \poloFantasmaCn{\rB}  &\qw     } 
\end{aligned}    \quad . 
 \end{equation*}
 The set of all pure states of system $\rA$ will be denoted  as $\Pure\cat C (\rI,\rA)$.  
 \end{definition}

Intuitively, a pure state is defined as an ``integral piece of information", which is independent of the surrounding context.      
From the definition it follows that  the product of two pure states is a pure state \cite{Chi14a,Chi14b}, namely  
\begin{align}
\alpha\otimes \beta  \in  \Pure\cat C  (\rI,\rA\otimes \rB) \qquad \forall \alpha  \in   \Pure\cat C(\rI,\rA) \, , \forall  \beta  \in  \Pure\cat C (\rI,\rB) \, .
\end{align}  

The definition of pure state can be extended in the obvious way to general gates, leading to the following  
\begin{definition}\label{def:purprocess} 
A gate $\map G  \in  \cat C(\rA,\rA')$ is \emph{pure}   iff it has only trivial extensions, that is, iff for every system $\rB\in |\cat C|$ and for every gate  $\map H\in  \cat C(\rA,\rA'\otimes\rB)$  one has the implication
\begin{equation*}
\begin{aligned}  \Qcircuit @C=.5em @R=0em @!R { &  \qw  \poloFantasmaCn{\rA}  &     \multigate{2}{\map H}    & \qw \poloFantasmaCn{\rA'}  &  \qw \\
  && \pureghost{\map H}   &&\\
 && \pureghost{\map H}    & \qw \poloFantasmaCn{\rB}  &  \measureD{\Tr}} 
\end{aligned}
 \quad    =  
\begin{aligned}  \Qcircuit @C=.5em @R=0em @!R 
{   & \qw  \poloFantasmaCn{\rA}    & \gate{  \map G }    &  \qw  \poloFantasmaCn{\rA'}  &\qw  } 
\end{aligned} 
   \qquad  \Longrightarrow \qquad      \exists  \beta\in  \cat C (\rI,\rB)  :  \quad  
   \begin{aligned}  \Qcircuit @C=.5em @R=0em @!R { &  \qw  \poloFantasmaCn{\rA}  &     \multigate{2}{\map H}    & \qw \poloFantasmaCn{\rA'}  &  \qw \\
  && \pureghost{\map H}   &&\\
 && \pureghost{\map H}    & \qw \poloFantasmaCn{\rB}  &  \qw} 
\end{aligned}
 \quad    =  
\begin{aligned}  \Qcircuit @C=.5em @R=0em @!R 
{   & \qw  \poloFantasmaCn{\rA}    & \gate{  \map G }    &  \qw  \poloFantasmaCn{\rA'}  &\qw  \\\\
&   &\prepareC{\beta}    &  \qw  \poloFantasmaCn{\rB}  &\qw    } 
\end{aligned} 
\quad . 
 \end{equation*}
 The set of all pure gates of type $\rA\to \rA'$ will be denoted  as $\Pure\cat C (\rA,\rA')$.  
 \end{definition}

\section{Programmability, distinguishability, and copiability}
We are now ready to introduce the three tasks that are  protagonists of this paper.

\subsection{Programmability}  

Consider the task of programming the operations performed by a machine using a set of instructions, encoded in the state of a physical system.  The idea can be formalized as follows:

\begin{definition}\label{def:prog}
Let $\set S  =  \{\rho_x \}_{  x\in  \set X }$ be a set of states of system $\rA$  and  let $\set G  =  \{   \map G_x  \}_{x\in \set X}$ be a set of gates of type $\rB\to \rB'$, with $\set X$ a suitable index set. 
 We say that the states    in $\set S$ \emph{program} the gates in $\set G$ iff  there exists a gate $\map W$, of type $\rA\otimes \rB \to \rB'$, such that 
\begin{equation}\label{prog}
\begin{aligned}
  \Qcircuit @C=1em @R=.7em @! R {  &\qw \poloFantasmaCn{\rB} &  \multigate{1} {\map W}  &   \qw \poloFantasmaCn{ \rB'} &\qw        & \quad &  =  & \quad  & \qw \poloFantasmaCn{\rB} &\gate{\map G_x}  & \qw \poloFantasmaCn{\rB'} &\qw    &&&&&\qquad \forall x \in \set X  \, .  \\
   \prepareC{\rho_x}&\qw \poloFantasmaCn{\rA} &  \ghost {\map W}  &   &  &&&&&&&&} 
   \end{aligned} 
\end{equation}
\end{definition}
\medskip 

In other words, the states in $\set S$ program the gates in $\set G$ if there exists a machine that can perform on demand every desired gate in $\set G$, controlled by a specific state in $\set S$.  Note that the gates in $\set G$ do not need  be reversible and, in general, their input and output can differ.  
For example, the input could be set to be the trivial system $\rB \equiv \rI$. In this case,  the gates in $\set G$ initialize system $\rB'$ in a given set of states $\{\beta_x\}_{x\in\set X}$ and Eq. (\ref{prog}) becomes 
\begin{equation}\label{progstate}
\begin{aligned}
  \Qcircuit @C=1em @R=.7em @! R {  \prepareC{\rho_x}&\qw \poloFantasmaCn{\rA} &  \gate {\map W}  &   \qw \poloFantasmaCn{ \rB'} &\qw  } 
   \end{aligned} 
  ~ =~ 
\begin{aligned}
  \Qcircuit @C=1em @R=.7em @! R {   \prepareC{\beta_x}& \qw \poloFantasmaCn{\rB'} & \qw}
\end{aligned}     \qquad \forall x \in \set X  \, .
   \end{equation}
 
\subsection{Distinguishability}  
In quantum theory, the density matrices in a given set $\{\rho_x\}_{x\in\set X}$  are (perfectly) distinguishable iff there exists a measurement, described by  operators $\{ P_x\}_{x\in\set X}$, satisfying the equation 
$$  \Tr [ P_x  \rho_y]  =  \delta_{xy}   \qquad \forall x,y\in \set X \, .$$ 
This definition cannot be exported to our abstract circuit model, however, because we do not  have  a notion of measurement.    Can we still make sense of the expression that some states are perfectly distinguishable?  

To answer this question, we should go at the root of  the operational meaning of distinguishability.  Operationally, the purpose of distinguishing states is to make decisions.  For example,  in a quantum state discrimination game the player would use the measurement $\{P_x\}_{x\in\set X}$ to decide which answer  $x\in \set X$ she should send to the referee.     One can also think of other types of games, where the player has a set of possible moves, described by a set  of  gates $\{\map G_x\}_{x\in\set X}$, and has  to choose one move depending on the information  contained in the state $\rho_x$.  
All these examples suggest that one can  \emph{identify} the ability to reliably distinguish states  with the ability to use them as programs for a desired set of operations. In the abstract circuit model, this intuition can be formalized as follows:  
\begin{definition}\label{def:dist}
Let $\set S  =  \{\rho_x \}_{  x\in  \set X }$ be a set of states of system $\rA$.   The states in $\set S$ are \emph{(perfectly) distinguishable} iff for every pair of systems $\rB, \rB' $ and for every indexed set of gates $\set G= \{  \map G_x\}_{x\in \set X}$  of type $\rB\to \rB'$ there exists a gate   $\map W_{\set G}:   \rA\otimes \rB  \to \rB'$ such that 
\begin{equation}\label{dist}
\begin{aligned}
  \Qcircuit @C=1em @R=.7em @! R {  &\qw \poloFantasmaCn{\rB} &  \multigate{1} {\map W_{\set G}}  &   \qw \poloFantasmaCn{ \rB'} &\qw   &  \quad &  =  & \quad  &  & \qw \poloFantasmaCn{\rB} &\gate{\map G_x}  & \qw \poloFantasmaCn{\rB'} &\qw  &&&&&    \qquad \forall x \in \set X  \, .\\
   \prepareC{\rho_x}&\qw \poloFantasmaCn{\rA} &  \ghost {\map W_{\set G}}  &   &  &&&&&&&&&} 
   \end{aligned} 
   \end{equation}
\end{definition}

\medskip In short, the states $\set S$ are perfectly distinguishable iff they can program every desired set of gates.   

\subsection{Distinguishability of the output implies distinguishability of the input}
An obvious consequence of definition \ref{def:dist} is the following: if the states in $\set S$ can be transformed into a set of distinguishable states, then they must be perfectly distinguishable:  
\begin{proposition} Let $\set S'   =  \{   \rho'_x\}_{x\in\set X}$ be a set of perfectly distinguishable states of system $\rA'$. 
 If   there exists  a gate  $\map A  :  \rA\to \rA'$ such that  
\begin{equation}\label{trans}
\begin{aligned}
  \Qcircuit @C=1em @R=.7em @! R {  \prepareC{\rho_x}&\qw \poloFantasmaCn{\rA} &  \gate {\map A}  &   \qw \poloFantasmaCn{ \rA'} &\qw  } 
   \end{aligned} 
  ~ =~ 
\begin{aligned}
  \Qcircuit @C=1em @R=.7em @! R {   \prepareC{\rho'_x}& \qw \poloFantasmaCn{\rA'} & \qw}
\end{aligned}     \qquad \forall x \in \set X  \, ,
   \end{equation}
   then the states $\{\rho_x\}_{x\in\set X}$ are perfectly distinguishable. 
   \end{proposition}
\Proof  Since the states  in $\set S'$ are perfectly distinguishable, for every indexed set of gates  $\set G= \{  \map G_x\}_{x\in\set X}$ there exists a gate $\map W_{\set G}'$ such that
\begin{equation*}
\begin{aligned}
  \Qcircuit @C=1em @R=.7em @! R {  &\qw \poloFantasmaCn{\rB} &  \multigate{1} {\map W'_{\set G}}  &   \qw \poloFantasmaCn{ \rB'} &\qw   &  \quad   &=  & \quad  &    \qw \poloFantasmaCn{\rB} &\gate{\map G_x}  & \qw \poloFantasmaCn{\rB'} &\qw     &&&&&   \qquad \forall x \in \set X  \, . \\
   \prepareC{\rho'_x}&\qw \poloFantasmaCn{\rA'} &  \ghost {\map W_{\set G}}  &   &} 
   \end{aligned} 
   \end{equation*}
Defining 
\begin{align*}
\begin{aligned}
  \Qcircuit @C=1em @R=.7em @! R {  &\qw \poloFantasmaCn{\rB} &  \multigate{1}{\map W_{\set G}}  &   \qw \poloFantasmaCn{ \rB'} &\qw  \\
   &\qw \poloFantasmaCn{\rA} &  \ghost {\map W_{\set G}}  &   &} 
   \end{aligned}  
   \quad :  =  \quad 
   \begin{aligned}
  \Qcircuit @C=1em @R=.7em @! R {   &\qw \poloFantasmaCn{\rB} &\qw&\qw  &  \multigate{1} {\map W'_{\set G}}  &   \qw \poloFantasmaCn{ \rB'} &\qw  \\
     & \qw \poloFantasmaCn{\rA}   &  \gate{\map A}   &\qw \poloFantasmaCn{\rA'} &  \ghost {\map W'_{\set G}}  &   &} 
   \end{aligned} 
\end{align*}
one obtains that the states in $\set S$ program the gates in $\set G$.  Since $\set G$ is arbitrary, this means that the states in $\set S$ are distinguishable. \qed 
\medskip 
  
 \subsection{Copiability}
Suppose that we are given a physical system $\rA$, with the promise that the system is in a state $\rho_x$ chosen from a set $\set S  =  \{\rho_x\}_{x\in\set X}$.   Thinking of the state as a piece of information, it is natural to ask whether it is possible to make copies of it.  In the abstract gate model,  we say that the states in $\set S$ are \emph{copiable}   iff there exists a gate $\map C  :  \rA\to \rA_1\otimes \rA_2$, with $\rA_1  \simeq \rA_2 \simeq \rA$, such that  
\begin{equation}\label{clon}
\begin{aligned}
  \Qcircuit @C=1em @R=.7em @! R {   \prepareC{\rho_x}  &\qw \poloFantasmaCn{\rA} &  \multigate{1} {\map C}  &   \qw \poloFantasmaCn{ \rA_1} &\qw  \\
  & &  \pureghost {\map C}  &     \qw \poloFantasmaCn{ \rA_2} &\qw  } 
   \end{aligned} 
  ~ =~ 
\begin{aligned}
  \Qcircuit @C=1em @R=.7em @! R { 
   \prepareC{\rho_x}  &\qw \poloFantasmaCn{\rA_1} &  \qw   \\
   \prepareC{\rho_x}  &\qw \poloFantasmaCn{\rA_2} &  \qw }
\end{aligned}   \qquad \forall x \in \set X \, .
   \end{equation}
%Note also that our general setting can be applied to pure theories, where all states are pure, and also  to mixed theories, where some states represent statistical mixtures.   classical and quantum computation as special cases. Also, it describes classical deterministic computation, in which all states of all systems are pure and they can be perfectly copied. 

\subsection{Distinguishability implies copiability}\label{subsect:distcop}

Suppose that the states in  $\set S$ are distinguishable.  Then, an immediate consequence of  definition  \ref{def:dist} is that they can be copied.  Indeed, we can choose the set  $\set G$ to consist of  gates  that initialize two  systems of type  $\rA$  in the states $ \{  \rho_x\otimes \rho_x\}_{x\in\set X}$.  Applying Eq. (\ref{dist}) to this particular set of gates we obtain  a gate $\map W_{\set G}$ such that
\begin{equation}
\begin{aligned}
  \Qcircuit @C=1em @R=.7em @! R {    \prepareC{\rho_x} &\qw \poloFantasmaCn{\rA} &  \multigate{1} {\map W_{\set G}}  &   \qw \poloFantasmaCn{ \rA_1} &\qw  \\
   &  &  \pureghost {\map W_{\set G}}  &    \qw \poloFantasmaCn{ \rA_2}   & \qw} 
   \end{aligned} 
  ~ =~ 
  \begin{aligned}
  \Qcircuit @C=1em @R=.7em @! R {   \prepareC{\rho_x}& \qw \poloFantasmaCn{\rA_1} & \qw  \\
   \prepareC{\rho_x}& \qw \poloFantasmaCn{\rA_2} & \qw}
\end{aligned}     \qquad \forall x \in \set X  \, .
   \end{equation}
 
A natural question is whether the converse is also true, namely whether copiable states are also distinguishable.   This result does \emph{not} follow  from the definitions given so far and requires some additional assumptions regarding distinguishability with multiple copies.      
These assumptions are spelt out in the next two sections. 

\section{Asymptotic distinguishability}

In this section we introduce a notion of  distinguishability in the asymptotic limit.  
   In order to do that,  we introduce  a topology on top of  the abstract circuit model. 
   
\subsection{Approximation of a gate}

The most  primitive notion of ``closeness" is the topological one.  In order to express the fact that two gates are ``close to one another" we introduce the following:  
\begin{definition}\label{def:topo}  
A \emph{topology for the circuit model} $\set C$ consists in the  assignment of a family of open subsets $\mathcal O_{\rA\to \rB}$ to every set of gates $\cat C (\rA ,\rB)$, in accordance with the following requirements
\begin{enumerate}
\item     $  \mathcal O_{\rA\to \rA'}   \times \mathcal O_{\rB\to \rB'}  \subseteq \mathcal O_{\rA\otimes \rB  \to \rA'\otimes \rB'}$ for arbitrary systems $\rA,\rA',\rB ,\rB'\in  |\cat C|$
\item    the insertion of a gate in a circuit is  continuous:  for all systems $\rA,\rB,\rC,\rD,\rR \in  |\cat C|$, 
  for every  pair of  gates $\map F  :  \rA \to \rB\otimes \rR$ and $\map H  :  \rC\otimes  \rR \to \rD$ and for  every open set $\set O \in  \mathcal O_{\rA\to \rD}$, the set    of gates  
$$   (  {\map F,\map H})^{-1}   \set O    :  = \left \{  \map G   \in  \cat C(\rB\to \rC)  \quad {\rm such~that}\quad    
\begin{aligned}
  \Qcircuit @C=1em @R=.7em @! R {     &\qw \poloFantasmaCn{\rA} &  \multigate{1} {\map F}  &   \qw \poloFantasmaCn{ \rB} &\gate{\map G}   &    \qw \poloFantasmaCn{ \rC}   &  \multigate{1}{\map H}   &     \qw \poloFantasmaCn{ \rD}   &\qw \\
  & &  \pureghost {\map F}  &    \qw & \qw \poloFantasmaCn{ \rR} &\qw & \ghost{\map H}   &  &     } 
   \end{aligned}  \in   \set O\right\} $$
is open. 
\end{enumerate}
\end{definition}

Using the above definition, we can express the fact that a sequence of gates converges to a specified gate. Precisely, the sequence $(\map G_n)_{n\in\mathbb N}  \subset  \cat C(\rA,\rB)$ converges to the gate $\map G$ iff for every open set $\set O  \in   \mathcal O_{\rA\to \rB}$ there exists an integer $N_{\set O}$ such that  one has
$$   \map G_n   \in  \set O   \qquad \forall n  > N_{\set O}  \, .$$
When this is the case, we write $\lim_{n\to \infty}\map G_n  =\map G$. 

In this paper, we assume the following 

\begin{axiom}\label{ax:compact}
For every pair of systems $\rA,\rB \in  |\cat C|$, we assume that the set of gates $\cat  C (\rA, \rB)$ is compact,~meaning that  for every sequence of gates $(   \map G_n)_{n\in\mathbb N}  \subset  \cat C(\rA,\rB)$ one can find a subsequence $(  \map G_{n_k})_{k  \in \mathbb N}$ and a gate $  \map G$ such that $  \lim_{k\to \infty}  \map G_{n_k}    = \map G.$  
\end{axiom}
%More generally, one can consider multiple sequences of gates   $\left( \map G_{x,n} \right)_{n\in\mathbb N}$, labelled by the index $x\in\set X$.   We say that the sequences \emph{converge}

\subsection{Approximate programmability}

The notion of approximation of gates allow us to discuss  approximate programmability:  
\begin{definition}\label{def:asymptoticprog}  
For every integer $n$, let $\rA_n$ be a system in $\cat C$ and let   $  \set S_n   :  =   \left\{  \rho_{x,n}\right\}$ be a set of states of  system $\rA_n$.       We say that the states in $\set S_n$   \emph{asymptotically program}  the gates  in  $\set G    =  \{\map G_x\}_{x\in\set X}$ iff  there exists a gate $  \map W_{\set G, n}   :   \rA_n\otimes  \rB\to \rB'$ such that  
\begin{equation}\label{asymptoticprog}
\lim_{n\to \infty}\begin{aligned}
  \Qcircuit @C=1em @R=.7em @! R {  &\qw \poloFantasmaCn{\rB} &  \multigate{1} {\map W_{\set G,n}}  &   \qw \poloFantasmaCn{ \rB'} &\qw  \\
   \prepareC{\rho_{x,n}}&\qw \poloFantasmaCn{\rA_n} &  \ghost {\map W_{\set G,n}}  &   &} 
   \end{aligned} 
     \quad =  \quad
\begin{aligned}
  \Qcircuit @C=1em @R=.7em @! R { & \qw \poloFantasmaCn{\rB} &\gate{\map G_x}  & \qw \poloFantasmaCn{\rB'} &\qw}
\end{aligned}   
   \end{equation}
 uniformly for every $x\in\set X$.   
 \end{definition}
 
 \medskip  
 
 %In the following we will be mostly concerned with finite sets of states, for which the requirement of uniform convergence is not needed. 
 Note that  we required that the convergence should be uniform in  $x$, because  this is  the appropriate requirement  when the set $\set X$ is infinite.

\subsection{Asymptotic  distinguishablity}

Using definition \ref{def:asymptoticprog} it is immediate to give a notion of asymptotic distinguishability of states:  
\begin{definition}For every integer $n$, let $\rA_n$ be a system in $\cat C$ and let   $  \set S_n   :  =   \left\{  \rho_{x,n}\right\}$ be a set of states of  system $\rA_n$.       We say that the states  $\set S_n$  are \emph{asymptotically distinguishable in the limit $n\to\infty$}  iff, for every pair of systems $\rB,\rB'\in  |\cat C|$ and every set of gates $ \set G  =  \{\map G_x\}_{x\in\set X}  \subset  \cat C(\rB,\rB')$, the states in $\set S_n$ asymptotically program the gates in $\map G$. 
\end{definition}

The notion of asymptotic distinguishability is very useful. Its usefulness is mostly due to the following proposition, which links distinguishability with its asymptotic version:

\begin{proposition}\label{prop:distdist}
Let $\set S  =  \{\rho_x\}_{x\in\set X}$ be set of states of system $\rA$ and, for every $n\in\mathbb N$,  let $\set S_n  =  \{  \rho_{x,n}\}_{x\in\set X}$ be a set of states of system $\rA_n$.        If the states in $\set S_n$ are asymptotically distinguishable and if there exists a gate $\map C_n:  \rA\to \rA_n$ such that 
\begin{align*}
\begin{aligned}
  \Qcircuit @C=1em @R=.7em @! R {  \prepareC{\rho_x}&\qw \poloFantasmaCn{\rA} &  \gate{{\map C}_n}  &    \qw \poloFantasmaCn{ \rA_n} &\qw  } 
   \end{aligned} 
  \quad = \quad 
\begin{aligned}
  \Qcircuit @C=1em @R=.7em @! R { \prepareC{\rho_{x,n} }& \qw \poloFantasmaCn{\rA_n} & \qw}
\end{aligned}     \qquad \forall x \in \set X  \, ,
\end{align*}
then the states in $\set S$ are distinguishable.  
\end{proposition}

\Proof 
Since the states in $\set S_n$ are asymptotically distinguishable, for every pair of systems $\rB,\rB'\in  |\cat C|$ and  every set of gates $\set G  = \{\map G_x\}_{x\in\set X}   \subset \cat C(\rB,\rB')$  there exists a gate $\map W_{\set G,n}$ such that   
\begin{equation*}
\lim_{n \to \infty}\begin{aligned}
  \Qcircuit @C=1em @R=.7em @! R {  &\qw \poloFantasmaCn{\rB} &  \multigate{1} {\map W_{\set G,n}}  &   \qw \poloFantasmaCn{ \rB'} &\qw  \\
   \prepareC{\rho_{x,n}}&\qw \poloFantasmaCn{\rA_n} &  \ghost {\map W_{\set G,n}}  &   &} 
   \end{aligned} 
  \quad = \quad
\begin{aligned}
  \Qcircuit @C=1em @R=.7em @! R { & \qw \poloFantasmaCn{\rB} &\gate{\map G_x}  & \qw \poloFantasmaCn{\rB'} &\qw}
\end{aligned}    \qquad \forall x\in\set X \, .
   \end{equation*}
Defining the gate
\begin{align*} 
\begin{aligned}
  \Qcircuit @C=1em @R=.7em @! R {  &\qw \poloFantasmaCn{\rB} &  \multigate{1} {\map Z_{\set G,n}}  &   \qw \poloFantasmaCn{ \rB'} &\qw  \\
   &\qw \poloFantasmaCn{\rA} &  \ghost {\map Z_{\set G,n}}  &   &} 
   \end{aligned}  
   \quad :  =  \quad 
   \begin{aligned}
  \Qcircuit @C=1em @R=.7em @! R {   &\qw \poloFantasmaCn{\rB} &\qw&\qw  &  \multigate{1} {\map W_{\set G,n}}  &   \qw \poloFantasmaCn{ \rB'} &\qw  \\
     & \qw \poloFantasmaCn{\rA}   &  \gate{\map C_n}   &\qw \poloFantasmaCn{\rA_n} &  \ghost {\map W_{\set G,n}}  &   &} 
   \end{aligned} 
\end{align*}
and combining the two equations above, one obtains  
\begin{equation*}
\lim_{n \to \infty}\begin{aligned}
  \Qcircuit @C=1em @R=.7em @! R {  &\qw \poloFantasmaCn{\rB} &  \multigate{1} {\map Z_{\set G,n}}  &   \qw \poloFantasmaCn{ \rB'} &\qw  \\
   \prepareC{\rho_{x}}&\qw \poloFantasmaCn{\rA} &  \ghost {\map Z_{\set G,n}}  &   &} 
   \end{aligned} 
  \quad  = \quad 
\begin{aligned}
  \Qcircuit @C=1em @R=.7em @! R { & \qw \poloFantasmaCn{\rB} &\gate{\map G_x}  & \qw \poloFantasmaCn{\rB'} &\qw}
\end{aligned}    \qquad \forall x\in\set X \, .
   \end{equation*}
Now, Axiom \ref{ax:compact} guarantees that there exists a subsequence  $(\map Z_{\set G,  n_k}  )_{k\in\mathbb N}$  and a gate $\map Z_{\set G}$ such that $  \lim_{k\to \infty}  \map Z_{\set G,n_k}     =   \map Z_{\set G}$.    Hence, one has  
\begin{align*}
\begin{aligned}
  \Qcircuit @C=1em @R=.7em @! R { & \qw \poloFantasmaCn{\rB} &\gate{\map G_x}  & \qw \poloFantasmaCn{\rB'} &\qw}
\end{aligned}       \quad   &=  \quad    \lim_{n\to \infty} 
\begin{aligned}
  \Qcircuit @C=1em @R=.7em @! R {  &\qw \poloFantasmaCn{\rB} &  \multigate{1} {\map Z_{\set G,n}}  &   \qw \poloFantasmaCn{ \rB'} &\qw  \\
   \prepareC{\rho_{x}}&\qw \poloFantasmaCn{\rA} &  \ghost {\map Z_{\set G,n}}  &   &} 
   \end{aligned} \\  \\
    & =  \quad 
   \lim_{k\to \infty} 
\begin{aligned}
  \Qcircuit @C=1em @R=.7em @! R {  &\qw \poloFantasmaCn{\rB} &  \multigate{1} {\map Z_{\set G,n_k}}  &   \qw \poloFantasmaCn{ \rB'} &\qw  \\
   \prepareC{\rho_{x}}&\qw \poloFantasmaCn{\rA} &  \ghost {\map Z_{\set G,n_k}}  &   &} 
   \end{aligned} \\  \\
   &  =
   \begin{aligned}
  \Qcircuit @C=1em @R=.7em @! R {  &\qw \poloFantasmaCn{\rB} &  \multigate{1} {\map Z_{\set G}}  &   \qw \poloFantasmaCn{ \rB'} &\qw  \\
   \prepareC{\rho_{x}}&\qw \poloFantasmaCn{\rA} &  \ghost {\map Z_{\set G}}  &   &} 
   \end{aligned}    \qquad \forall x\in\set X \, ,
   \end{align*}
where the last equality used the fact that the composition of gates is continuous (cf. item 2 of definition \ref{def:topo}). In conclusion, we proved that every set of gates $\set G$ can be programmed by the states in  $\set S$.  By definition \ref{def:dist}, this means that the states in $\set S $ are distinguishable. \qed

\section{Asymptotic i.i.d. distinguishability}  

Suppose that we are given a large number of identical systems of type $\rA$, prepared in the i.i.d. state $\rho_x^{\otimes N}$,  $x\in\set X$.  Intuitively, if  the states $\set S$ are distinct from one another, then it should be possible to find out the value of $x$ with vanishing error.    This is the case in quantum theory, where one can perform quantum state tomography and identify the state $\rho_x$ up to an error that vanishes when the number of copies  goes to infinity.
Of course, the tomography  argument requires one to have a notion of measurement, which has not been introduced in the framework so far.     In order to express the intuitive property of asymptotic  i.i.d. distinguishability, one has two alternatives:  The first alternative is to introduce measurements and probabilities.  When this is done,  one can \emph{prove} a theorem stating that  the probability of error in the identification of the label $x$  vanishes in the limit $N\to \infty$  \cite{Lei07,Chi10}.    The second alternative is to \emph{assume} asymptotic i.i.d. distinguishability as an axiom. 
Here we follow this route:
\begin{axiom}%[Asymptotic i.i.d. distinguishability]
\label{ax:iid}
For every system $\rA\in  |\cat  C|$ and for every set of distinct states of $\rA$, say $S=\{ \rho_x\}_{x\in\set X} $, the i.i.d states $ \{  \rho_x^{\otimes n} \}_{x\in\set X} $   are asymptotically distinguishable.    
\end{axiom}

In the next sections we will explore the consequences of this requirement.

\section{Distinguishability and generation of side information}

Distinguishability is closely related with another operational task, which consists in generating some additional piece of data from a given state.  Formally, the task is defined as follows: 
\begin{definition}
Let $\set S  = \{  \rho_x\}_{x\in\set X}$ be a set of distinct states of system $\rA$.  We say that the gate   $\map C:  \rA\to \rA\otimes \rE$ \emph{generates side information}   for   the   states in  $\set S$  iff  there exists a set of states of system $\rE$, say  $\{\eta_x\}_{x\in\set X}$, such that 
\begin{equation}\label{side}
\begin{aligned}
  \Qcircuit @C=1em @R=.7em @! R {   \prepareC{\rho_x}  &\qw \poloFantasmaCn{\rA} &  \multigate{1} {\map C}  &   \qw \poloFantasmaCn{ \rA} &\qw  \\
  & &  \pureghost {\map C}  &     \qw \poloFantasmaCn{ \rE} &\qw  } 
   \end{aligned} 
  ~ =~ 
\begin{aligned}
  \Qcircuit @C=1em @R=.7em @! R { 
   \prepareC{\rho_x}  &\qw \poloFantasmaCn{\rA} &  \qw   \\
   \prepareC{\eta_x}  &\qw \poloFantasmaCn{\rE} &  \qw }
\end{aligned}   \qquad \forall x \in \set X \, ,
   \end{equation}
and at least two states $\eta_{x_0}$ and $\eta_{x_1}$ are distinct.    Moreover, we say that the gate $\map C$  \emph{generates faithful side information}  iff the states $\{\eta_x\}_{x\in\set X}$ are all distinct.  
\end{definition}

One example of process that generates faithful side information is copying:  in this particular case, one has $\rE  =  \rA$ and $\eta_x  =  \rho_x$ for every $x\in\set X$.  

We now show that only distinguishable states allow one to  generate  \emph{faithful} side information:   
\begin{proposition}\label{prop:faithside}
The following are equivalent: 
\begin{enumerate}
\item  the states $\set S$ are  distinguishable
\item there exists a gate that generates faithful side information for $\set S$. 
\end{enumerate}
\end{proposition} 

\Proof  Clearly, if the states are distinguishable, one can use them to program the preparation of the states $\{\rho_x\otimes \eta_x\}_{x\in\set X}$   for every desired set  of states $\{\eta_x\}_{x\in\set X} $  of every desired system $\rE$.     Conversely, suppose that there exists a gate  $\map C$ that generates side information for the states in $\set S$.    By applying the gate  $\map C$  twice, one obtains  
\begin{equation*}
\begin{aligned}
  \Qcircuit @C=1em @R=.7em @! R {   \prepareC{\rho_x}  &\qw \poloFantasmaCn{\rA} &  \multigate{2} {\map C}  &   \qw \poloFantasmaCn{ \rA} &\multigate{1}{\map C}    &\qw \poloFantasmaCn{\rA} &  \qw   \\
  & &  \pureghost {\map C}  &      &      \pureghost{\map C}      &\qw \poloFantasmaCn{\rE} &\qw \\
  & &  \pureghost {\map C}  &     \qw \poloFantasmaCn{ \rE} &\qw   &\qw &\qw } 
   \end{aligned} 
  ~ =~ 
\begin{aligned}
  \Qcircuit @C=1em @R=.7em @! R { 
   \prepareC{\rho_x}  &\qw \poloFantasmaCn{\rA} &  \qw   \\
   \prepareC{\eta_x}  &\qw \poloFantasmaCn{\rE} &  \qw \\
    \prepareC{\eta_x}  &\qw \poloFantasmaCn{\rE} &  \qw }
\end{aligned}   \qquad \forall x \in \set X \, .
  \end{equation*}
More generally, applying $\map C$  for $n$ times one obtains a gate $\map C_n:  \rA\to \rA\otimes \rE^{\otimes n}$ such that 
\begin{align*}
\begin{aligned}
  \Qcircuit @C=1em @R=.7em @! R {  \prepareC{\rho_x}&\qw \poloFantasmaCn{\rA} &  \multigate{1}{\map C_n}  &   \qw \poloFantasmaCn{ \rA} &\qw \\
  & &  \pureghost{\map C_n}  &   \qw \poloFantasmaCn{ \rE^{\otimes n}} &\qw  } 
   \end{aligned} 
  \quad = \quad 
\begin{aligned}
  \Qcircuit @C=1em @R=.7em @! R { \prepareC{\, \rho_x^{\phantom{\otimes ~ }}}  &\qw \poloFantasmaCn{\rA} &  \qw \\   \prepareC{\eta_x^{\otimes n}}& \qw \poloFantasmaCn{\rE^{\otimes n}} & \qw}
\end{aligned}     \qquad \forall x \in \set X  \, ,
\end{align*}
Discarding the output system $\rA$, one obtains the gate   
\begin{align*}
\begin{aligned}
  \Qcircuit @C=1em @R=.7em @! R {  &\qw \poloFantasmaCn{\rA} &  \gate{\widetilde{\map C}_n} &   \qw \poloFantasmaCn{ \rE^{\otimes n}} &\qw  } 
   \end{aligned} 
   := 
\begin{aligned}
  \Qcircuit @C=1em @R=.7em @! R {  &\qw \poloFantasmaCn{\rA} &  \multigate{1}{\map C_n}  &   \qw \poloFantasmaCn{ \rA} &  \measureD{\Tr}   \\
  & &  \pureghost{\map C_n}  &   \qw \poloFantasmaCn{ \rE^{\otimes n}} &
 \qw } 
   \end{aligned} \quad ,
   \end{align*}
  which satisfies  
\begin{align*}
\begin{aligned}
  \Qcircuit @C=1em @R=.7em @! R {  \prepareC{\rho_x}&\qw \poloFantasmaCn{\rA} &  \gate{\widetilde{\map C}_n}  &    \qw \poloFantasmaCn{ \rE^{\otimes n}} &\qw  } 
   \end{aligned} 
  \quad = \quad 
\begin{aligned}
  \Qcircuit @C=1em @R=.7em @! R { \prepareC{\eta_x^{\otimes n}}& \qw \poloFantasmaCn{\rE^{\otimes n}} & \qw}
\end{aligned}     \qquad \forall x \in \set X  \, ,
\end{align*}
due to the normalization condition of Eq. (\ref{statenorm}).  
Now, by hypothesis the states $\{\eta_x\}_{x\in\set X} $ are distinct.  Hence, by Axiom \ref{ax:iid} the states in the set $\set S_n:  =\left\{\eta_x^{\otimes n} \right\}_{x\in\set X} $ are asymptotically distinguishable. Proposition \ref{prop:distdist} then guarantees that the states in $\set S$  are distinguishable.  
\qed  

\medskip

When the side information is not faithful,  the situation is slightly more diversified. 
In analogy with Refs.  \cite{Lov79,Dua13,Chi13} we   define  the \emph{confusability graph}   of a set of states   $\set S  =  \{\rho_x\}_{x\in\set X}$ as the graph where
  \begin{enumerate}
  \item  the vertices are the elements of $\set X$
  \item two vertices $x$ and $y$ are adjacent iff the corresponding states $\rho_x$ and $\rho_y$ are not distinguishable.  
  \end{enumerate} 
Let us denote by $\{\set X_k\}_{k=1}^K$ the connected components of the confusability graph.  We then have the following
\begin{proposition}
If the gate $\map C :  \rA\to   \rA\otimes \rE$ generates side information for the set $\set S$, then  for every connected component $\set X_k$   there exists a state $\eta_k\in\cat C (\rI,\rE)$   such that 
\begin{equation*}
\begin{aligned}
  \Qcircuit @C=1em @R=.7em @! R {   \prepareC{\rho_x}  &\qw \poloFantasmaCn{\rA} &  \multigate{1} {\map C}  &   \qw \poloFantasmaCn{ \rA} &\qw  \\
  & &  \pureghost {\map C}  &     \qw \poloFantasmaCn{ \rE} &\qw  } 
   \end{aligned} 
  ~ =~ 
\begin{aligned}
  \Qcircuit @C=1em @R=.7em @! R { 
   \prepareC{\rho_x}  &\qw \poloFantasmaCn{\rA} &  \qw   &&& \qquad \forall x \in \set X_k \, .    \\
   \prepareC{\eta_k}  &\qw \poloFantasmaCn{\rE} &  \qw }
\end{aligned}  
   \end{equation*}
\end{proposition}

 \Proof   By definition, the fact that the gate $\map C $ generates side information amounts to the condition 
 \begin{equation*}
\begin{aligned}
  \Qcircuit @C=1em @R=.7em @! R {   \prepareC{\rho_x}  &\qw \poloFantasmaCn{\rA} &  \multigate{1} {\map C}  &   \qw \poloFantasmaCn{ \rA} &\qw  \\
  & &  \pureghost {\map C}  &     \qw \poloFantasmaCn{ \rE} &\qw  } 
   \end{aligned} 
  ~ =~ 
\begin{aligned}
  \Qcircuit @C=1em @R=.7em @! R { 
   \prepareC{\rho_x}  &\qw \poloFantasmaCn{\rA} &  \qw   &&& \qquad \forall x \in \set X \, .    \\
   \prepareC{\eta_x}  &\qw \poloFantasmaCn{\rE} &  \qw }
\end{aligned}  
   \end{equation*}
   We have to show that $\eta_x$  does not depend on the particular element $x$, but only to the connected component it belongs to.   This is easily done thanks to proposition \ref{prop:faithside}, which guarantees that  if $x$ and $y$ are connected, then  $\eta_x  =  \eta_y$.  \qed

\section{Copiability-Distinguishability  Equivalence} 

In the previous section we saw that only distinguishable states can generate faithful side information.   This fact  implies a fundamental  equivalence between copiability and distinguishability.  

\begin{proposition}%[Cloning-Distinguishability Equivalence]
Let $\set S\subset \cat  C(\rI,\rA)$ be a finite set of distinct states.    The states $\set S$ are copiable if and only if  they are distinguishable. 
\end{proposition} 

\Proof  We already saw in subsection \ref{subsect:distcop} that distinguishable states are copiable.   
  Conversely, suppose that the states in $\set S$ are copiable with a gate $\map C$, as in Eq. (\ref{clon}).    By definition, the gate $\map C$ generates faithful side information.  Hence, by proposition \ref{prop:faithside} the states $\set S$ must be perfectly distinguishable. \qed   
  
\section{Cryptographic No Information Without Disturbance}  

At the qualitative level, the security of the  many   quantum key distribution protocols (such as e.g. \cite{Ben84}) is based on the fact that when  a quantum system is prepared in a pure   state chosen from a set of two (or more) non-orthogonal states,  an eavesdropper cannot extract any information about the state of the system without changing the state of the system.    We refer to this feature as the Cryptographic No Information Without Disturbance property.  An iconic demonstration of this working principle  is the B92 protocol \cite{Ben92}, which employs the transmission of just two non-orthogonal states.   
 It is then natural to wonder under which conditions this feature can be reproduced in a general process theory, other than quantum theory.  
 
Here we show that, if one accepts the definitions  given in this paper, the Cryptographic No Information Without Disturbance is a logical implication, valid in arbitrary theories.  
We model the process of  extracting  information from  system $\rA$ as a gate $\map G$ of type   $ \rA\to \rA\otimes \rE$, where $\rE$  is the system held by the eavesdropper.  For the information encoded in the states $\alpha_0$ and $\alpha_1$, the condition of no disturbance is   
\begin{align}\label{nodist}
\begin{aligned}
  \Qcircuit @C=1em @R=.7em @! R {   \prepareC{\alpha_x}  &\qw \poloFantasmaCn{\rA} &  \multigate{1} {\map G}  &   \qw \poloFantasmaCn{ \rA} &\qw      &       &   =     &      & \prepareC{\alpha_x}  &\qw \poloFantasmaCn{\rA}   &\qw        &&&&  &&&  \forall  x  \in  \{0,1\} \, ,   \\
  & &  \pureghost {\map G}  &     \qw \poloFantasmaCn{ \rE} &\measureD {\Tr}  &&& &&&&& &&&&&} 
   \end{aligned}
   \end{align}
   meaning that the  marginal state of system $\rA$ is not affected by the presence of the  gate  $\map G$.  
On the other hand, the condition that  the gate $\map G$ extracts information from the input is  that the marginal state of system $\rE$ depends on the input label $x$, namely
\begin{align}\label{info}
\begin{aligned}
  \Qcircuit @C=1em @R=.7em @! R {   \prepareC{\alpha_0}  &\qw \poloFantasmaCn{\rA} &  \multigate{1} {\map G}  &   \qw \poloFantasmaCn{ \rA} & \measureD {\Tr} \\
  & &  \pureghost {\map G}  &     \qw \poloFantasmaCn{ \rE} &\qw    } 
   \end{aligned}
   \quad   \not  =  \quad  
   \begin{aligned}
  \Qcircuit @C=1em @R=.7em @! R {   \prepareC{\alpha_1}  &\qw \poloFantasmaCn{\rA} &  \multigate{1} {\map G}  &   \qw \poloFantasmaCn{ \rA} & \measureD {\Tr} \\
  & &  \pureghost {\map G}  &     \qw \poloFantasmaCn{ \rE} &\qw    } 
   \end{aligned} \quad .
   \end{align}
 With these  definitions we have the following
\begin{proposition}
Let   $  \alpha_0$ and $\alpha_1$ be two pure states of system $\rA$.  If the two states not distinguishable, then no information can be extracted without disturbance, that is,  Eqs.  (\ref{nodist}) and (\ref{info}) cannot be jointly satisfied.  
\end{proposition}   

\Proof Suppose that the no disturbance condition of Eq. (\ref{nodist}) is satisfied. Then,  by definition of pure state (definition \ref{def:purstate}), one must have 
\begin{align*}
\begin{aligned}
  \Qcircuit @C=1em @R=.7em @! R {   \prepareC{\alpha_x}  &\qw \poloFantasmaCn{\rA} &  \multigate{1} {\map G}  &   \qw \poloFantasmaCn{ \rA} &\qw        \\
  & &  \pureghost {\map G}  &     \qw \poloFantasmaCn{ \rE} &  \qw } 
   \end{aligned}  
   \quad =  \quad 
\begin{aligned}
  \Qcircuit @C=1em @R=.7em @! R {
     \prepareC{\alpha_x}  &\qw \poloFantasmaCn{\rA} &    \qw       &&&&&&&  \qquad \forall x\in\{0,1\}    \\
     \prepareC{\eta_x}  &\qw \poloFantasmaCn{\rE} &    \qw      } 
   \end{aligned}  
   \end{align*}
   for some (not necessarily pure) states $\eta_0$ and $\eta_1$.    Now, if   $\eta_0\not  =  \eta_1$, then the gate $\map G$ generates faithful side information for the states $\{\alpha_0,\alpha_1\}$.  By proposition \ref{prop:faithside}, this implies that $\alpha_0$ and $\alpha_1$ are distinguishable.  Since by hypothesis  the states $\alpha_0$ and $\alpha_1$ are not distinguishable by hypothesis, we conclude that Eq. (\ref{info}) cannot be satisfied.  \qed    

\section{Summary and outlook}  

In this paper we formulated the  basic notion of distinguishability   without reference to probabilities.  Our definition, formulated in an abstract circuit model,  expresses the intuitive fact that  two pieces of information are distinguishable if they can be used  as instructions to program every two desired operations.   
We then examined the relation between distinguishability and copiability, which required us to enrich the circuit model with  a topology.      Thanks to this enrichment, we have been able to discuss a   notion of asymptotic distinguishability and to require as an axiom that a state can be identified with arbitrary precision from a sufficiently large number of copies.

Once the above notions have been put into place, we established a number of relations among the notions of distinguishability, copiability, and programming.   
First of all, we showed that the  states in a given set  are distinguishable if and only if one can generate some side information from them.   From this basic result we derived two facts: \emph{i)} the equivalence between distinguishability and copiability, and \emph{ii)} the Cryptographic No Information Without Disturbance.

The present work is part of a larger program  of categorification of the framework of operational-probabilistic theories \cite{Chi10}, which aims at reducing the probabilistic part of the framework  at the advantage of the operational one.   In this spirit,  an interesting open question for future research is whether the notion of state broadcasting and the no-broadcasting theorem for general probabilistic theories \cite{Lei07} can be imported to the probability-free scenario.  

\medskip  

{\bf Acknowledgements.}    This work was supported   by the National Basic Research Program of China
(973) 2011CBA00300 (2011CBA00301) and  by the National
Natural Science Foundation of China (Grants 11450110096,
61033001, 61061130540), by the 1000 Youth Fellowship
Program of China, by the Foundational Questions Institute (Grants FQXi-RFP3-1325). GC acknowledges the hospitality of Perimeter Institute for Theoretical Physics.  Research at Perimeter Institute for Theoretical Physics is supported in part by the Government of Canada through NSERC and by the Province of Ontario through MRI.

%\nocite{*}
%\bibliographystyle{eptcs}


\begin{thebibliography}{50}

\bibitem[Abr04]{Abr04} Abramsky, S. and Coecke, B.: A categorical semantics of quantum protocols, 
 in \emph{Proc. of the 19th Annual IEEE Symposium on Logic in
Computer Science}, pp. 415-425 (2004).
 \bibitem[Abr08]{Abr08}   Abramsky, S. and Coecke: Categorical quantum mechanics,  \emph{Handbook of quantum logic and
 quantum structures},   Engesser, K. ,
 Gabbay,  D. M. and 
 Lehmann, D. eds.,  Elsevier, pp. 261-324 (2008).
\bibitem[Abr09]{Abr09}  Abramsky, S.:  No Cloning in Categorical Quantum Mechanics, 
\emph{Semantic Techniques in Quantum Computation},   Gay, S.   and  Mackie,  I. eds, pp. 1-28, Cambridge University Press (2010).
\bibitem[Awo10]{Awo10} Awodey, S.:   \emph{Category theory.}
 Oxford University Press, 2010.    
 	\bibitem[Bar07]{Bar07}  Barrett, J.:  
	Information processing in generalized probabilistic theories,  
	\emph{Phys. Rev. A}, {\bf 75}, 032304  (2007).   
\bibitem[Bar12]{Bar12}  Barnum, H.,  Barrett,  J.,  Leifer, M. and  Wilce, A.:
Teleportation in general probabilistic theories,  
\emph{Proc. of Symposia in Applied Mathematics}, {\bf 71}, pp. 25-48 (2012).
\bibitem[Bar11]{Bar11}   Barnum, H.  and  Wilce, A.: Information processing in convex operational theories, 
\emph{Electronic Notes in Theoretical Computer Science} {\bf 270}(1), pp. 3-15  (2011).  
\bibitem[Ben84]{Ben84}   Bennett, C. H.  and  Brassard, G.:  Quantum cryptography: Public key distribution and coin tossing, in \emph{Proceedings of IEEE International Conference on Computers, Systems and Signal Processing}, {\bf 175}, p. 8 New York (1984).  
 \bibitem[Ben92]{Ben92}   Bennet, C. H.:  Quantum cryptography using any two nonorthogonal states,
 \emph{Phys. Rev. Lett.}, {\bf  68}, 3121 (1992).   
\bibitem[Chi10]{Chi10}   Chiribella,  G.,  D'Ariano,  G. M. and  Perinotti,  P.: Probabilistic theories with purification, 
\emph{ Phys. Rev. A}, {\bf 81}, p. 062348  (2010).	
\bibitem[Chi11]{Chi11}  Chiribella,  G.,  D'Ariano,  G. M. and  Perinotti,  P.:  Informational derivation of quantum theory, \emph{Phys. Rev. A},  {\bf  84}, p. 012311  (2011).
\bibitem[Chi13]{Chi13}  Chiribella, G. and Yang, Y.:  Confusability graphs for symmetric sets of quantum states,
in \emph{Symmetries and Groups in Contemporary Physics},  Bai, C.-M., Gazeau, J. P.,  and  Ge, M.- L. eds., Nankai Series in Pure, Applied Mathematics and Theoretical Physics, 11 (2013).
\bibitem[Chi14a]{Chi14a}  Chiribella, G.:  Categorical purification, \\  
\emph{http://www.cs.ox.ac.uk/CQM2014/programme/Giulio.pdf} (2014). 
\bibitem[Chi14b]{Chi14b}  Chiribella, G.:  Purity without probability, manuscript in preparation (2014). 
\bibitem[Coe06]{Coe06}  Coecke,  B.: Kindergarten quantum mechanics: lecture notes, 
   in \emph{ AIP Conf. Proc.},  {\bf  810},   pp. 81-98  (2006).
   \bibitem[Coe08]{Coe08}   Coecke, B.:     Axiomatic description of mixed states from Selinger's CPM-construction, 
   \emph{Electronic Notes in Theoretical Computer Science},  {\bf 210}, pp. 3?13  (2008).
   \bibitem[Coe10]{Coe10}  Coecke, B.: Quantum picturalism, \emph{ Contemporary physics}, {\bf  51}(1),  pp. 59?83 (2010).
    \bibitem[Har01]{Har01}  Hardy, L.:
     Quantum theory from five reasonable axioms,
     	\emph{arXiv:quant-ph/0101012}  (2001).  
	\bibitem[Har11]{Har11}   Hardy, L.: Reformulating and reconstructing quantum theory,  arXiv preprint arXiv:1104.2066   (2011).
	\bibitem[Hor11]{Hor11} Horseman, C.:  Quantum picturalism for topological cluster-state computing, 
   \emph{ New J. Phys.},   {\bf 13},  095011 (2011).
   \bibitem[Per10]{Per10} Coecke, B. and  Perdrix, S.: Environment and classical channels in categorical quantum mechanics,  in \emph{Computer Science Logic}, Springer, pp. 230?244  (2010).
   \bibitem[Coe12]{Coe12}   Coecke,  B.,  Duncan,  R.,  Kissinger, A. and  Wang, Q.:  Strong Complementarity and Non-locality in Categorical Quantum Mechanics, 
 	\emph{  Proc.  of the  27th Annual IEEE/ACM Symposium on Logic in Computer Science}, pp. 245-254 (2012).   
	\bibitem[Coe14]{Coe14}  Coecke, B.: Terminality implies non-signalling, \emph{arXiv preprint arXiv:1405.3681}  (2014).
\bibitem[Dak11]{Dak11}  Dakic,  B.  and Bruckner, C.:   Quantum Theory and Beyond: Is Entanglement Special?,  
  in \emph{Deep Beauty: Understanding the Quantum World through Mathematical Innovation},  Halvorson,  H. ed.,   pp.
365?392,    Cambridge University Press, Cambridge, 2011. 
 \bibitem[DAr06]{DAr06}   D'Ariano, G. M.: 
On the missing axiom of quantum mechanics,
\emph{AIP Conf. Proc.},  {\bf 810},   pp. 114-130 (2006).  
\bibitem[DAr10]{DAr10}   D'Ariano, G. M.: {Probabilistic theories: what is special about quantum mechanics}, in \emph{Philosophy of quantum information and entanglement}, Bokulich, A. and Jaeger, G. eds., p.  85  (2010).
  \bibitem[Dua13]{Dua13}  Duan,  R.,  Severini,  S., and Winter,  A.: Zero-error communication via quantum
channels, non-commutative graphs and a quantum Lov\'asz $\theta$-function,  \emph{IEEE Trans. Inf. Theory},   {\bf 59}(2), 1164-1174 (2013).  
\bibitem[Lal12]{Lal12}  Coecke, B. and Lal, R. A.: Causal categories: relativistically interacting processes,   \emph{Found. Phys.}, {\bf43}(4), 458-501 (2012).
\bibitem[Lei07]{Lei07}   Barnum, H., Barrett,  J.,  Leifer, M. and  Wilce,  A.: 
Generalized no-broadcasting theorem,  
\emph{ Phys. Rev. Lett. }, {\bf 99}(24), p. 240501 (2007).
\bibitem[Lov79]{Lov79}   Lov\'asz,  L.:    On the Shannon capacity of a graph, 
\emph{ IEEE Trans. Inform. Th.} IT-{\bf 25}, 1 (1979).
\bibitem[Kel80]{Kel80}  Kelly,  G. M. and  Laplaza, M. L.: Coherence for compact closed categories, \emph{Journal of Pure
and Applied Algebra},  {\bf 19}, pp. 193-213 (1980).
 \bibitem[Mas11]{Mas11}  Masanes,   L. and   M\"uller,  M. P.:    A derivation of quantum theory from physical requirements,  \emph{   New J. Phys.},  {\bf 13}, 063001 (2011).  
 \bibitem[Mas13]{Mas13}     Masanes, L.,   M\"uller,  M. P. , Augusiak,  R., P\'erez-Garc\'ia,  D.:     A digital approach to quantum theory, \emph{   Proc. of the National Academy
of Sciences}, {\bf 110}, 16373 (2013).
\bibitem[Mer07]{Mer07} Mermin, N. D.:   \emph{Quantum Computer Science: an Introduction.}
 Cambridge University Press, 2007.    
\bibitem[Nie01]{Nie01} Nielsen, M. A.  and Chuang, I. L.:   \emph{Quantum Computation and Quantum Information.}
 Cambridge University Press, 2001.    
\bibitem[Ran14]{Ran14}  Ranchin,  A. and Coecke,  B.:   Complete set of circuit equations for Stabilizer Quantum Mechanics, \emph{Phys. Rev. A} {\bf  90}, 012109 (2014). 
 \bibitem[Spe07]{Spe07}  Spekkens, R. W.: Evidence for the epistemic view of quantum states: A toy theory,  
 \emph{ Phys. 
Rev. A}, {\bf 75}(3), p. 032110  (2007).
 \bibitem[Sel11]{Sel11}   Selinger, P.: A survey of graphical languages for monoidal categories, 
  in \emph{New structures for
physics}, Coecke, B. ed., Springer, pp. 289-355 (2011). 
\bibitem[Wil13]{Wil13} Wilde, M.:   \emph{Quantum Information Theory.}
 Cambridge University Press, 2007.    
  \end{thebibliography}
\end{document}